\documentclass[prb,aps,twocolumn,showpacs,preprintnumbers,amsmath,amssymb]{revtex4-1}
\usepackage{graphicx}
\usepackage{dcolumn}
\usepackage{bm}
\usepackage{multirow}
\usepackage{tabularx,booktabs}
\usepackage{ulem}
\begin{document}


\title{ Generation Efficiencies for Propagating Modes in a Supersolid } 
\author{ Matthew R. Sears} 
\author{ Wayne M. Saslow} 
\email{wsaslow@tamu.edu}
\affiliation{ Department of Physics, Texas A\&M University, College Station, TX 77843-4242}
\date{\today}

\begin{abstract}
Using Andreev and Lifshitz's supersolid hydrodynamics, we obtain the propagating longitudinal modes at non-zero applied pressure $P_{a}$ (necessary for solid $^4$He), and their generation efficiencies by heaters and transducers.  For small $P_{a}$, a solid develops an internal pressure $P \sim P_{a}^2$.  This theory has stress contributions both from the lattice and an internal pressure $P$.  Because both types of stress are included, the normal mode analysis differs from previous works.   Not surprisingly, transducers are significantly more efficient at producing elastic waves and heaters are significantly more efficient at producing fourth sound waves.  We take the system to be isotropic, which should apply to systems that are glassy or consist of many crystallites; the results should also apply, at least qualitatively, to single-crystal hcp $^4$He.  
\end{abstract}



\maketitle
\section{Introduction}
In 1969 Andreev and Lifshitz developed a theory of supersolids.\cite{AL69}  Although the microscopic physical description was for flow of vacancies, the macroscopic equations did not depend on vacancies in an essential fashion.  At about the same time Thouless\cite{Thouless69} and Chester\cite{Chester70} both suggested the possibility of superflow in a solid by vacancies.  In addition, Leggett\cite{Leggett70} pointed out the possibility of Non-Classical Rotational Inertia (NCRI) associated with quantum-mechanical flow via a superfluid velocity (a phase gradient) opposite the local velocity of the rotating lattice.  

Since the observation of NCRI by Kim and Chan,\cite{KC1,KC2} a number of laboratories have reproduced their work.\cite{RR1,Shira1,Koj1,Penzev,RR2,RR3,Lin07,ClarkWestChan07}  (For reviews that emphasize experiment, see Refs.~\onlinecite{BalCau08} and \onlinecite{ChanScience08}.) Were NCRI the sole criterion for superflow of solids, there would be strong reason to accept that such superflow has been observed.  However, a supersolid should also have other properties, including a fourth sound-like mode, as predicted by Andreev and Lifshitz, and modified elastic waves with higher velocities, since the superfluid mass does  not participate in the motion.  (We remind the reader that a fourth sound mode in superfluid $^4$He occurs only when the normal fluid is entrained by a porous medium; in the present case the lattice serves as the porous medium.)  Neither a fourth sound mode nor velocity shifts have been observed.\cite{Aoki2007,Aoki2008,Kwon10}  However, a stiffened shear response {\it is} observed,\cite{Day06, DayBeamishNature07} although not enough to explain the observed NCRI.\cite{ChanScience08}  Note also recent work indicating that supersolidity in $^4$He can only occur below 55~mK.\cite{ShevDayBeam10}  

As a guide to experiments to observe the fourth sound mode,\cite{Kwon10} the present work calculates various quantities relevant to its observation, such as the relative efficiencies of a transducer and a heater in producing both longitudinal elastic waves and fourth sound waves.  It also 
considers the effect of a non-zero applied pressure $P_a$; to solidify $^4$He, even near $T=0$, requires $P_a \gtrsim 25$~bar.  To our knowledge, previous works have not included the effect of $P_{a}$.  

Although we believe that vacancies are essential to a microscopic understanding of superflow in solids, in the hydrodynamic theory they play no fundamental role, other than as an additional variable largely tied to diffusion.  Indeed, we believe that the hydrodynamic theory is more likely to describe a supersolid related to the NCRI effect proposed by Leggett than to vacancy superflow.  

Most of the present work assumes that the system is isotropic.  One effect this has is that the superfluid density, which properly is a second rank tensor $\tensor{\rho_s}$, is proportional to the unit matrix, so we take $\tensor{\rho_s}\approx\tensor{1}\rho_{s}$.\cite{SasJolad,GalliSas}  We then write the superfluid fraction as 
\begin{equation}
f_{s}=\frac{\rho_{s}}{\rho}, \label{fs}
\end{equation}
where $\rho_{s}$ is the superfluid density and $\rho$ is the total (mass) density.  $f_s$ is unity in a superfluid at low temperatures.  However, in putative supersolid $^4$He, the measured NCRI fraction, which if due to superflow should be equated to $f_{s}$, is never greater than about 0.2.  The {\it effective} normal fraction $f_{n}$ thus has the curious property of being not less than 0.8, although at $T=0$ there are no excitations to destroy the superflow.  We have previously noted this difficulty,\cite{Saslow05} and proposed that the lattice be given a mass fraction $f_{L}$, in addition to a contribution $f_{n}^{ex}$ due to excitations, so that $1=f_{s}+f_{n}^{ex}+f_{L}$.  This permits, at $T=0$, no excitations (so $f_{n}^{ex}=0$) but $f_{s}<1$.  In this viewpoint, the lattice velocity is identified with $\dot{u}_{i}$, where $u_{i}$ is the lattice displacement, and the effective normal fluid fraction $f_n$ is the sum of $f_{L}$ and a part $f_{n}^{ex}$ due to excitations: $f_{n}=f_{n}^{ex}+f_{L}$.


It is known that the more annealed (and thus more crystalline) the sample of $^4$He, the smaller the NCRI fraction.\cite{RR2}  Likewise it is known that the more quenched the sample, the larger the NCRI fraction.\cite{RR3}  Hence the supersolidity is more likely to occur for less crystalline samples, which might be either glassy\cite{NussinovBalatsky} or consist of a large number of small crystallites.\cite{SasakiBalibar}  In both of these cases an acoustic probe is likely to take a rotational average, thus making the system behave more like an isotropic system than a crystal.  
Therefore we consider systems whose macroscopic properties are isotropic.  If a pure crystal of hcp $^4$He were to be supersolid, then the results we obtain will be only an approximation; nevertheless they will be a useful guide for experiment.

We also note that we are working in the linear regime, where the disturbances produced by a heater or transducer are expected to be only a small perturbation, as is assumed in all theories of this sort.  It is possible that solid $^4$He is ultrasensitive to temperature or to stress (e.g. if it is a glass, perhaps the atoms can be driven off their sites by a transducer).  Nevertheless, there should always be a linear regime; Ref.~\onlinecite{Kwon10} notes that their membrane for producing putative fourth sound produced strains much below the critical value.

The present work is intended to be self-contained, although we do not derive the equations of motion, which may be most explicitly obtained from Sect.IV of Ref.~\onlinecite{Saslow77},\cite{GalileanNote} nor do we derive the $P_a$ dependence of various quantities, which is done explicitly in 
Ref.~\onlinecite{SearsSasOrdinary}.  
 
Sections~\ref{ThermoSection} and \ref{EqsOfMotionSection} present the thermodynamics and equations of motion, respectively.  Section~\ref{FreqModeSection} studies the eigenfrequency and eigenmodes for each of the longitudinal propagating modes.  Section~\ref{StressTemp} finds, for each mode, the stress and temperature response in terms of the normal fluid velocity.  Section~\ref{ModeDetectionSection} finds and discusses the efficiency of generating each propagating mode by a transducer and a heater.  Section~\ref{Conclusion} provides a brief summary of the results.  

One of the features of the hydrodynamic theory of supersolids is that it contains stress due to both  internal pressure $P$ and the lattice, in order to permit the system properties to continuously transform into those of a superfluid.  This means that under an applied pressure $P_{a}$ the pressure and the lattice each take up a part of it.  We believe that use of both a pressure and a lattice stress is needed not merely for solid $^4$He but for other solids as well, particularly those under pressure or with point defects that are not in equilibrium.\cite{SearsSasOrdinary}  Appendix~\ref{PPaAppendix} discusses the relationship between the internal pressure $P$ and the applied pressure $P_a$, which we estimate using experimental data and results from Ref.~\onlinecite{SearsSasOrdinary}.   
Appendix~\ref{ApproximationAppendix} discusses the relative sizes of velocities and strains in a crystal under an applied pressure.  Appendix~\ref{dTAppendix} finds the relative size of two thermodynamic derivatives of temperature $T$ that appear in the generation efficiencies.  

\section{Thermodynamics}
\label{ThermoSection}
The thermodynamic equations for a supersolid are given in terms of the energy density $\epsilon$, entropy density $s$, unsymmetrized strain $w_{ij}=\partial_{i}u_{j}$, mass density $\rho$, superfluid velocity $\vec{v}_{s}$, and momentum density 
\begin{align}
\vec{g} = \rho_{n}\vec{v}_{n} + \rho_{s}\vec{v}_{s}, \label{g}
\end{align}
and their thermodynamically conjugate quantities.\cite{AL69, Saslow77}  Here $\vec{v}_n$ is the normal fluid velocity.  Specifically,
\begin{align}
d\epsilon &= Tds+\lambda_{ik}dw_{ik}+ \mu d\rho +\vec{v}_n\cdot d\vec{g} +
\vec{j}_s \cdot d\vec{v}_s, \label{depsilon}\\ 
\epsilon &= -P+Ts+\lambda_{ik}w_{ik}+\mu \rho +\vec{v}_n \cdot\vec{g}+ \vec{j}_s
\cdot \vec{v}_s, \label{epsilon}\\
0 &= -dP+sdT+w_{ik}d\lambda_{ik}+\rho d\mu +\vec{g}\cdot d\vec{v}_n + \vec{v}_s
\cdot d\vec{j}_s. \label{GibbsDuhem}
\end{align}
Here the thermodynamically conjugate quantities are temperature $T$, (unsymmetrized) elastic tensor density $\lambda_{ik}$ (with units of pressure $P$), chemical potential $\mu$ (with units of velocity squared), normal fluid velocity $\vec{v}_{n}$, and 
\begin{equation}
 \vec{j}_s = \vec{g} - \rho \vec{v}_n = \rho_s (\vec{v}_s - \vec{v}_n). \label{js}
\end{equation}
Note that $\vec{j}_{s}$ has units of momentum density but is invariant under Galilean boosts.  That is, if both $\vec{v}_{n}$ and $\vec{v}_{s}$ are boosted by $\delta \vec{v}$, $\vec{j}_{s}$ does not change. 

We find it convenient to define 
\begin{equation}
 \vec{j}_n \equiv \rho \vec{v}_n, \label{jn}\\
\end{equation}
so that 
\begin{equation}
\vec{g} = {\vec{j}}_n +  {\vec{j}}_s.
\label{gjnjs}
\end{equation}
Unlike $\vec{j}_{s}$, the quantity $\vec{j}_{n}$ is a momentum density both in units {\it and} in its properties under Galilean boosts;  under a boost by $\delta \vec{v}$, both $\vec{g}$ and $\vec{j}_n$ are boosted by $\rho(\delta \vec{v})$. 



\section{Hydrodynamic Equations}
\label{EqsOfMotionSection}
Eq.~\eqref{depsilon} shows that there are five independent thermodynamic variables.  Two of them are scalars ($s$ and $\rho$), one is a tensor ($w_{ik}=\partial_{i}u_{k}$) and two are vectors ($\vec{g}$ and $\vec{v}_{s}$).  In developing the hydrodynamic equations we will employ the first three, but we will use the two vectors $\vec{j}_{n}$ and $\vec{j}_{s}$ in place of $\vec{g}$ and $\vec{v}_{s}$.  For an  ordinary solid, where $\vec{v}_{s}$ does not appear, it is convenient to use the variables $\sigma \equiv s/\rho$, $\rho$, $w_{ik}$, and $\vec{g}$, since $\sigma$ decouples from the other variables.  Such decoupling does not occur for the supersolid.  Note that one could also use the scalar variables $T$ and $\rho$, or $T$ and $\mu$, or $s$ and $\mu$.   


Unless otherwise specified, thermodynamic derivatives with respect to $\rho$, $s$, or $w_{ik}$ are taken with the other two variables held constant. 

We consider small amplitude excitations of the form $\exp{[i (\vec{k} \cdot \vec{r} - \omega t)]}$, where the wavevector $\vec{k}$ is taken to be known.  Then, with primes denoting deviations from equilibrium, in the absence of damping the equations of motion are given by\cite{AL69, Saslow77}
\begin{align}
&\dot{\rho}' + \partial_i g_i' = 0, \label{rho1}\\
&\dot{g}_i' - \partial_k \sigma_{ik}' = 0, \quad \sigma_{ik} \equiv \lambda_{ik} - P \delta_{ik}, \label{g1}\\
&\dot{v}_{si}' + \partial_i \mu' = 0, \label{vs1}\\
&\dot{s}' + s \partial_i {v_n}_i' = 0, \label{s1}\\
&\dot{u}_i' - {v_n}_i' = 0. \label{u1}
\end{align}
Rather than the stress tensor $\sigma_{ik}$, the momentum flux $\Pi_{ik}= - \sigma_{ik} + \rho_s {v_s}_i {v_s}_k + \rho_n {v_n}_i {v_n}_k \approx - \sigma_{ik}$ has also been employed,\cite{AL69,Khalatnikov} as well as $g_i = j_i$.\cite{AL69,Khalatnikov,LLFluid} 

In terms of the thermodynamic variables $\rho$ and $s$, we have
\begin{align}
&\mu' = \frac{\partial \mu}{\partial \rho} \rho' + \frac{\partial \mu}{\partial s} s' + \frac{\partial \mu}{\partial w_{jl}} w_{jl}',\label{muprime}\\
&\sigma_{ik}' = \frac{\partial \sigma_{ik}}{\partial \rho} \rho' + \frac{\partial \sigma_{ik}}{\partial s} s' + \frac{\partial \sigma_{ik}}{\partial w_{jl}} w_{jl}'.
\label{sigmaprime}
\end{align}
The equations of motion \eqref{rho1}, \eqref{s1}, and \eqref{u1} directly give
\begin{align}
&\rho' = \frac{k_i g_i'}{\omega}, \label{rho2}\\
&s' = s\frac{k_i  {v_n}_i'}{\omega}, \label{s2}\\
&w_{ij}' = i k_i u_j' = - \frac{k_i {v_n}_j'}{\omega}. \label{u2}
\end{align}
The other two equations of motion, \eqref{g1} and \eqref{vs1}, can now be written in terms of ${v_s'}_i$ and ${v_n'}_i$, or, equivalently, ${j_s'}_i$ and ${j_n'}_i$.

From \eqref{sigmaprime}-\eqref{u2}, momentum conservation \eqref{g1} gives
\begin{align}
\omega g_i' =& -k_k \sigma_{ik}' \notag\\ 
=& -\frac{k_k k_l}{\omega} \frac{\partial \sigma_{ik}}{\partial \rho} g_l' - \frac{k_k k_l}{\omega} s \frac{\partial \sigma_{ik}}{\partial s} {v_n}_l' + \frac{k_k k_j}{\omega} \frac{\partial \sigma_{ik}}{\partial w_{jl}} {v_n}_l'. \label{g2}
\end{align}
We now rearrange to use the variables ${j_n'}$ and ${j_s'}$.  From \eqref{jn}-\eqref{gjnjs}, multiplying \eqref{g2} by $\omega$ gives
\begin{align}
&0 = \left[ \omega^2 \delta_{il} +  k_k k_l \frac{\partial \sigma_{ik}}{\partial \rho}\right] {j_s}_l' + \left[ \omega^2 \delta_{il} + \left(k_k k_l \frac{\partial \sigma_{ik}}{\partial \rho} \right.  \right.  \notag\\
& \qquad \qquad \qquad  \left.  \left.+ k_k k_l \frac{s}{\rho} \frac{\partial \sigma_{ik}}{\partial s} - k_k k_j \frac{1}{\rho} \frac{\partial \sigma_{ik}}{\partial w_{jl}} \right) \right] {j_n}_l'. 
\label{g4}
\end{align}

Likewise, from \eqref{muprime} and \eqref{rho2}-\eqref{u2}, the superfluid equation of motion \eqref{vs1} gives
\begin{align}
\omega {v_s}_i' =& k_i \mu' 
= \frac{k_i k_l}{\omega} \frac{\partial \mu}{\partial \rho} g_l' + \frac{k_i k_l}{\omega} s \frac{\partial \mu}{\partial s} {v_n}_l' - \frac{k_i k_j}{\omega} \frac{\partial \mu}{\partial w_{jl}} {v_n}_l'. 
\label{vs2}
\end{align}
From \eqref{jn}-\eqref{gjnjs} and $\rho_s {v_s}_i' = {j_s}_i' - (\rho_s/\rho) {j_n}_i'$, multiplying \eqref{vs2} by $\rho_s \omega$ and rearranging gives
\begin{align}
&0 = \left[\omega^2 \delta_{il} - k_i k_l f_s \rho \frac{\partial \mu}{\partial \rho} \right] {j_s}_l' + f_s\bigg[ \omega^2 \delta_{il}  \notag\\
& \qquad   - \rho \left(k_i k_l  \frac{\partial \mu}{\partial \rho} +  k_i k_l  \frac{s}{\rho} \frac{\partial \mu}{\partial s} - k_i k_j  \frac{1}{\rho} \frac{\partial \mu}{\partial w_{jl}} \right) \bigg]{j_n}_l'.
\label{vs4}
\end{align}

Equations \eqref{g4} and \eqref{vs4} yield the normal mode frequencies and their eigenvectors (the ratio of the responses of the normal and superfluid currents).  In what follows we consider only an isotropic solid.  The effect this constraint has on \eqref{g4} and \eqref{vs4} is that the second-rank tensors are all proportional to the unit tensor, and the term $k_k k_j{\rho}^{-1}({\partial \sigma_{ik}}/{\partial w_{jl}})$ in \eqref{g4} contains two terms, one proportional to $\delta_{il}$ and one proportional to $k_{i}k_{l}$.  

Taking the dot product of these equations with $k_{i}$ then gives two equations in the unknowns $\omega^{2}$, $\vec{k}\cdot\vec{j}'_{s}$, and $\vec{k}\cdot\vec{j}'_{n}$.  This yields $\omega^{2}$ and the ratio $\vec{k}\cdot\vec{j}'_{s}/\vec{k}\cdot\vec{j}'_{n}$.  In addition, taking the cross-product of \eqref{vs4} with $k_{i}$ gives, since $\vec{k}\times\vec{v}_{s}=\vec{0}$, identically zero.  Further, taking the cross-product of \eqref{g4} with $k_{i}$ gives an equation having terms proportional to $\omega^{2}$ and $k^{2}$, both multiplying $\vec{k}\times\vec{v}_{n}$.  

There are two ways to solve the resulting equations for $\vec{k}\cdot\vec{j}'_{s}$, $\vec{k}\cdot\vec{j}'_{n}$, and $\vec{k}\times\vec{v}_{n}$.  One solution is to take $\vec{k}\times\vec{v}_{n}=\vec{0}$ and $k_{l}j'_{sl}\ne0$, $k_{l}j'_{nl}\ne0$ (purely longitudinal modes), with the frequencies determined by the two equations in the unknowns $\omega^{2}$, $\vec{k}\cdot\vec{j}'_{s}$, and $\vec{k}\cdot\vec{j}'_{n}$.  Hence this set of modes is purely longitudinal.  The other solution is to take $\vec{k}\times\vec{v}_{n}\ne\vec{0}$ and $\vec{k}\cdot\vec{j}'_{s}=0$, $\vec{k}\cdot\vec{j}'_{n}=0$ (purely transverse modes), with the frequencies determined by the the cross-product of \eqref{g4} with $k_{i}$.  $\vec{v}_{s}$ does not participate in the transverse modes, so their mass weighting involves only $\rho/\rho_{n}$, and their frequencies squared should be higher than in the normal solid by $\rho/\rho_{n}$.  To our knowledge such an effect has not been observed.\cite{AokiFootnote}  


\section{Longitudinal Eigenfrequencies and Eigenmodes}
\label{FreqModeSection}
Recall that, unless otherwise specified, thermodynamic derivatives with respect to $\rho$, $s$, or $w_{ik}$ are taken with the other two variables held constant.

\subsection{Some Properties and Definitions}
We now compute the quantity $\partial\sigma_{ik}/\partial\rho$, which appears in \eqref{g4}.  We take the strain response of a solid to $P_a$ to be isotropic (i.e., $w_{ik}^{(0)} \sim \delta_{ik} w_{ll}^{(0)}$, where the superscript $(0)$ denotes the static value).  Recall that $w_{ik}$ is unsymmetrized; here we take only the static part, due to $P_a$, to be symmetric, as does Ref.~\onlinecite{LLElasticity}.  Then, by Ref.~\onlinecite{LLElasticity},
\begin{align}
\lambda_{ik}^{(0)} = \left(K - \frac{2}{3} \mu_V \right)\delta_{ik} w_{ll}^{(0)}  + \mu_V \left(w_{ik}^{(0)} + w_{ki}^{(0)} \right)\sim \delta_{ik} w_{ll}^{(0)},
\label{lambda}
\end{align}
so that we can write
\begin{equation}
\frac{\partial \lambda_{ik}}{\partial \rho} \equiv \frac{\partial \lambda}{\partial \rho}\delta_{ik}.
\label{dlambda/drho}
\end{equation}
Here, $K$ and $\mu_V$ are the respective bulk and shear moduli, with units of $P$; $\mu_V$ is completely distinct from $\mu$.  Eq.~\eqref{dlambda/drho} is also employed in Ref.~\onlinecite{SearsSasOrdinary}, although there $\sigma=s/\rho$ is held constant rather than $s$.  At $T \approx 0$, the difference should be negligible. 
 Thus we can write 
\begin{align}
\frac{\partial \sigma_{ik}}{\partial \rho} = \frac{\partial \lambda_{ik}}{\partial \rho} - \delta_{ik}\frac{\partial P}{\partial \rho} = \left[\frac{\partial \lambda}{\partial \rho} - \frac{\partial P}{\partial \rho}\right] \delta_{ik} \equiv \frac{\partial \widetilde{\sigma}}{\partial \rho} \delta_{ik},
\label{dsigmadrho0}
\end{align}
where we use $\widetilde{\sigma}$ to distinguish a stress (with the same units as $\sigma_{ik}$) from $\sigma=s/\rho$; $\widetilde{\sigma}$ and $\sigma$ are not related.  Note that $\partial \widetilde{\sigma}/\partial \rho$ is not a true derivative, merely a definition; further, we do not here define a $\widetilde{\sigma}$.

We now compute the quantity $\partial\sigma_{ik}/\partial w_{jl}$, which also appears in \eqref{g4}. Since Ref.~\onlinecite{SearsSasOrdinary} shows that $(\partial P/\partial w_{ik}) \sim w_{ik}^{(0)} \sim w_{ll}^{(0)} \delta_{ik}$, we can write 
\begin{align}
\frac{\partial P}{\partial w_{ik}} \equiv \frac{\partial P}{\partial w}\delta_{ik}.
\label{dPdw}
\end{align}
$\partial P/\partial w$ is evaluated in Ref.~\onlinecite{SearsSasOrdinary}, and is given in Appendix~\ref{ApproximationAppendix}.  We also use the definitions
\begin{align}
\frac{\partial \lambda}{\partial w} \equiv K+ \frac{4}{3}\mu_V,\qquad \frac{\partial \widetilde{\sigma}}{\partial w} \equiv \frac{\partial \lambda}{\partial w} - \frac{\partial P}{\partial w},
\end{align}
As above, $\partial \widetilde{\sigma}/\partial w$ is not a true derivative, merely a definition. 
Eqs.~\eqref{lambda} and \eqref{dPdw} then give
\begin{align}
\frac{\partial \sigma_{ik}}{\partial w_{jl}} = \left(\frac{\partial \widetilde{\sigma}}{\partial w} - 2 \mu_V \right) \delta_{ik} \delta_{jl} + \mu_V \delta_{il} \delta_{jk} + \mu_V \delta_{ij}\delta_{kl}.
\label{dsigmadwGeneral1}
\end{align}
Thus, 
\begin{align}
k_k k_j \frac{\partial \sigma_{ik}}{\partial w_{jl}} {j_n'}_l= \left(\frac{\partial \widetilde{\sigma}}{\partial w} - \mu_V \right) k_i (\vec{k}\cdot\vec{j}'_{n}) + \mu_V k^2 {j_n'}_i.
\label{dsigmadwGeneral2}
\end{align}
For $\vec{k}\cdot\vec{j}'_{n}\neq 0$ and $\vec{k} \times \vec{j}_n =0$ (the longitudinal case), $k_i (\vec{k}\cdot\vec{j}'_{n}) = k^2 {j_n'}_i$ so that \eqref{dsigmadwGeneral2} gives
\begin{align}
k_k k_j \frac{\partial \sigma_{ik}}{\partial w_{jl}} {j_n'}_l= \frac{\partial \widetilde{\sigma}}{\partial w} k^2 \mu_V {j_n'}_i.
\label{dsigmadw}
\end{align}

It is convenient to define the velocities $c_0$ and $c_1$, which satisfy
\begin{align}
c_0^2 \equiv& \rho \frac{\partial \mu}{\partial \rho}, \\
c_1^2 \equiv& -\frac{\partial \widetilde{\sigma}}{\partial \rho}  + \frac{1}{\rho}\frac{\partial \widetilde{\sigma}}{\partial w} . 
\label{c1DEF}
\end{align}
If $\sigma$, rather than $s$, were held constant, then $c_0$ would be the sound velocity in an ordinary fluid, and $c_1$ would be the velocity of sound in an ordinary solid with no superflow.\cite{SearsSasOrdinary}  Using the Gibbs-Duhem relation \eqref{GibbsDuhem} 
and neglecting thermal expansion and terms second order in velocities gives
\begin{align}
\frac{\partial P}{\partial \rho} = c_0^2 + w_{jl} \frac{\partial \lambda_{jl}}{\partial \rho}.
\end{align}
Then, eq.~\eqref{dsigmadrho0} gives 
\begin{align}
\frac{\partial \widetilde{\sigma}}{\partial \rho}  \approx&  \left(1-w_{ll}^{(0)} \right) \frac{\partial \lambda}{\partial \rho} - c_0^2 .
 \label{dsigmadrho}
\end{align}
In the following we use either \eqref{dsigmadrho0} or \eqref{dsigmadrho}, depending on convenience.

\subsection{Reducing the Equations of Motion}

{\bf Momentum Equation \eqref{g4}:}
We take $s (\partial \sigma_{ik}/\partial s) \rightarrow 0$, which should be a reasonable approximation for solid $^4$He at low temperatures, both because $s \rightarrow 0$ as $T \rightarrow 0$, and because $K$ and $\mu_V$ (and therefore $\lambda_{ik}$ at constant $w_{jl}$) should be nearly independent of $s$.  Substituting \eqref{dsigmadrho0} and \eqref{dsigmadw} into \eqref{g4} and using \eqref{c1DEF} and \eqref{dsigmadrho} then gives, for a purely longitudinal mode,
\begin{align}
0 =& \left[ \omega^2  -   \left(\widetilde{c}^2 +w_{ll}^{(0)} \frac{\partial \lambda}{\partial \rho} \right)k^2 \right] {j_s}' + \left[ \omega^2  -  c_1^2 k^2 \right] {j_n}';
\label{g4-2}
\end{align}
here we define, to simplify the equations, 
\begin{align}
\widetilde{c}^2 \equiv c_0^2 - \frac{\partial \lambda}{\partial \rho}.
\label{gammaDEF}
\end{align} 
Appendix~\ref{ApproximationAppendix} finds that $|\partial \lambda/\partial \rho| \gg c_0^2$, so that $\widetilde{c}^2 \approx  - \partial \lambda/\partial \rho$.  It also finds that $\widetilde{c}^2$ is expected to be positive, and first order in $P_a/K$.  Further, it shows that for $P_a \ll K$ we have $c_1^2 \gg \widetilde{c}^2 \gg c_0^2$. 

{\bf Superfluid Equation \eqref{vs4}:}
A Maxwell relation that follows from \eqref{depsilon}, combined with \eqref{dlambda/drho}, gives
\begin{align}
\frac{\partial \mu}{\partial w_{jl}} = \frac{\partial \lambda_{jl}}{\partial \rho} = \delta_{jl} \frac{\partial \lambda}{\partial \rho}.
\end{align}
Then, neglecting $s (\partial \mu/\partial s) = s (\partial T/\partial \rho) \sim T^4$, and taking the mode to be purely longitudinal, eq.~\eqref{vs4} gives
\begin{align}
&0 = \left[\omega^2  -  f_s c_0^2 k^2 \right] {j_s}' + f_s\left[ \omega^2  - k^2 \widetilde{c}^2  \right]{j_n}'.
\label{vs4-2}
\end{align}

We use \eqref{g4-2} and \eqref{vs4-2} first to find the longitudinal mode frequencies, then to find the superfluid-to-normal ratios of current density and velocity in each longitudinal mode.  For $f_{s}\rightarrow0$, eq.~\eqref{vs4-2} gives either $\omega^{2}=f_s c^{2}_{0}k^{2}$ (fourth sound) or $j'_{s}=0$ (no superflow).  In the latter case, substitution into \eqref{g4-2} then gives $\omega^{2}=c^{2}_{1}k^{2}$ (first sound).

\subsection{Longitudinal Mode Frequencies}
Eqs. \eqref{g4-2} and \eqref{vs4-2} yield
\begin{align}
0 =&\omega^4 (1-f_s) - \omega^2 k^2 \left[ c_1^2 + f_s c_0^2 - f_s \left(2\widetilde{c}^2 + w_{ll}^{(0)} \frac{\partial \lambda}{\partial \rho} \right) \right] \notag\\
&+k^4 f_s \left[ c_1^2 c_0^2 - \widetilde{c}^2\left(\widetilde{c}^2 + w_{ll}^{(0)} \frac{\partial \lambda}{\partial \rho} \right)\right].
\label{Freq1-2}
\end{align}

Solving \eqref{Freq1-2} to first order in $f_s$ gives 
\begin{align}
&\frac{\omega_1^2}{k^2} \equiv \frac{\omega_+^2}{k^2} 
= c_1^2 + f_s \left[c_1^2 - 2 \widetilde{c}^2 + \frac{\widetilde{c}^4}{c_1^2}  + w_{ll}^{(0)}\frac{\partial \lambda}{\partial \rho}\left( \frac{\widetilde{c}^2}{c_1^2} -1\right) \right] ,
\label{FirstSoundFreq-2}
\end{align}
and
\begin{align}
\frac{\omega_4^2}{k^2} \equiv \frac{\omega_-^2}{k^2} 
= f_s\left( c_0^2 - \frac{\widetilde{c}^4}{c_1^2}- w_{ll}^{(0)} \frac{\partial \lambda}{\partial \rho} \frac{\widetilde{c}^2}{c_1^2} \right) \equiv f_s \widetilde{c}_0^2.
\label{FourthSoundFreq-2}
\end{align}
In the limit where $c_0^2 \ll {\partial \lambda}/{\partial \rho} $ and $w_{ll}^{(0)} \ll 1$ (i.e., $P_a \ll K$), 
\begin{align}
\widetilde{c}_0^2 \approx c_0^2 -  \frac{\widetilde{c}^4}{c_1^2}.
\label{barc4}
\end{align}
%
Appendix~\ref{ApproximationAppendix} finds that both terms on the right-hand-side of \eqref{barc4} are second order in $P_a/K$.  Further, it shows that for $P_a \ll K$ we have $c_1^2 \gg \widetilde{c}^2 \gg \widetilde{c}_0^2$.

\subsection{Longitudinal Mode Structure -- Currents and Velocities}
We now find the ratios of the normal fluid and superfluid response for both longitudinal modes.  These ratios will be used to calculate, for each mode, the response to the stress and temperature produced by transducers and by heaters.  We employ
\begin{align}
\frac{v_s'}{v_n'} = \frac{\rho}{\rho_s} \frac{\rho_s (v_s' - v_n')}{\rho v_n'} + 1 = \frac{1}{f_s}  \frac{j_s'}{j_n'} + 1.
\label{vsTOvn-2}
\end{align}

The ratios $j'_{s}/j'_{n}$ for each mode can in principle be obtained from the normal mode frequencies and either of \eqref{g4-2} or \eqref{vs4-2}.  

\subsubsection{First Sound Mode Structure}
\label{1stSoundStructure}
From \eqref{vs4-2}, with the subscript $1$ denoting first sound, 
\begin{equation}
\frac{{j_s'}_1}{{j_n'}_1} = - f_s \frac{ \frac{\omega_1^2}{k^2} -  \widetilde{c}^2}{\frac{\omega_1^2}{k^2} -  f_s c_0^2}.
\end{equation}
Substituting $\omega_{1}^{2}$ from \eqref{FirstSoundFreq-2}, accurate to zeroth order in $f_{s}$, gives a ratio accurate to first order in $f_{s}$: 
\begin{align}
\frac{{j_s'}_1}{{j_n'}_1} \approx  -f_s \left(1 - \frac{\widetilde{c}^2}{c_1^2} \right).
\label{js1TOjn1-2}
\end{align}
Then, using \eqref{vsTOvn-2}, the ratio of superfluid velocity to normal velocity for first sound is
\begin{align}
\frac{{v_s'}_1}{{v_n'}_1} \approx& \frac{\widetilde{c}^2}{c_1^2}.
\label{vs1TOvn1-2}
\end{align}
Appendix~\ref{ApproximationAppendix} shows that $c_1^2 \gg \widetilde{c}^2$, so ${v_n'}_1 \gg {v_s'}_1$.

\subsubsection{Fourth Sound Mode Structure}
\label{4thSoundStructure}
From \eqref{g4-2}, with subscript 4 denoting fourth sound, 
\begin{align}
\frac{{j_s'}_4}{{j_n'}_4} = - \frac{\frac{\omega_4^2}{k^2} - c_1^2}{\frac{\omega_4^2}{k^2} - \left[ \widetilde{c}^2 + w_{ll}^{(0)} \frac{\partial \lambda}{\partial \rho} \right]}.
\label{js4TOjn4generic-2}
\end{align}
With $\omega_{4}^{2}\sim f_{s}$, for $f_s \ll 1$ 
\begin{align}
\frac{{j_s'}_4}{{j_n'}_4} \approx&  - \frac{c_1^2}{ \widetilde{c}^2 + w_{ll}^{(0)} \frac{\partial \lambda}{\partial \rho}} 
.
\label{js4TOjn4-2a}
\end{align}
Appendix~\ref{ApproximationAppendix} shows that if $P_a/K \ll 1$,  then $w_{ll}^{(0)} \ll 1$ and $c_0^2 \ll \partial \lambda/\partial \rho$.  Thus, eq.~\eqref{gammaDEF} gives $\widetilde{c}^2 \approx -\partial \lambda/\partial \rho \gg w_{ll}^{(0)} \partial \lambda/\partial \rho$.  Then, 
  \begin{align}
\frac{{j_s'}_4}{{j_n'}_4} \approx&  - \frac{c_1^2}{ \widetilde{c}^2 } 
.
\label{js4TOjn4-2}
\end{align}
Then, using \eqref{vsTOvn-2}, the ratio of superfluid velocity to normal velocity for fourth sound is, to lowest order in $f_s$,
\begin{align}
\frac{{v_s'}_4}{{v_n'}_4} \approx - \frac{c_1^2}{f_s \widetilde{c}^2 } .
\label{vs4TOvn4-2}
\end{align}
Appendix~\ref{ApproximationAppendix} shows that $c_1^2 \gg \widetilde{c}^2$, so ${v_s'}_4 \gg {v_n'}_4$.

\section{Longitudinal Modes -- Stress and Temperature Responses}
\label{StressTemp}
We now calculate the deviations from equilibrium of the longitudinal stress and temperature produced by a transducer and by a heater.  We consider that only the $\sigma'_{11}$ component of the stress is generated.  For notational simplicity we employ $\hat{\sigma}' \equiv\sigma'_{11}$; recall that $\sigma$ is reserved for the entropy/mass. 

\subsection{Stress}
\label{Stress}
Conservation of momentum \eqref{g1} yields
\begin{align}
\hat{\sigma}' = -\frac{\omega g'}{k} = -\frac{\omega}{k} (j_s' + j_n') =  -\frac{\omega}{k} \left(\frac{j_s'}{j_n'} + 1 \right) \rho v_n' .
\label{sigma1}
\end{align}
Substituting the ratio ${j_s'}/{j_n'}$ from \eqref{js1TOjn1-2} and \eqref{js4TOjn4-2} and $\omega_{1,4}=+ c_{1,4} k$ from \eqref{FirstSoundFreq-2} and \eqref{FourthSoundFreq-2} into \eqref{sigma1} gives the stress associated with each mode.  For $  f_s \ll 1$, 
\begin{align}
\hat{\sigma}'_1 \approx& -c_1\left[1 - f_s \left(1-\frac{\widetilde{c}^2}{c_1^2}\right)\right]\rho {v'_n}_1\approx -\rho c_1 {v'_n}_1, \label{sigmaFirst}\\
\hat{\sigma}'_4 \approx& f_s^{\frac{1}{2}}  \frac{c_1^2}{\widetilde{c}^2 } \rho \widetilde{c}_0   {v_n'}_4 = f_s^{\frac{1}{2}}  \frac{c_1^2}{\widetilde{c}^2 } \rho \widetilde{c}_0 \frac{{v_n'}_4}{{v_n'}_1}  {v_n'}_1.
\label{sigmaFourth}
\end{align} 
where we have used $c_1^2 \gg \widetilde{c}^2$ (see Appendix~\ref{ApproximationAppendix}).  

The total stress deviation therefore is 
\begin{align}
\hat{\sigma}'= \hat{\sigma}'_1 + \hat{\sigma}'_4 \approx  -c_1 \rho {v'_n}_1 \left[1 - f_s^{\frac{1}{2}} \widetilde{c}_0  \frac{c_1}{\widetilde{c}^2 }   \frac{{v_n'}_4}{{v_n'}_1} \right].
\label{sigmatot}
\end{align} 
The ratio ${v_n'}_4/{v_n'}_1$ depends on the mode generator, to be discussed in the next section. 

\subsection{Temperature}
\label{Temperature}
The temperature deviation is less straightforward to obtain because it is a function of the variables $s$, $\rho$, and $w_{jl}$:
\begin{align}
T' \approx \frac{\partial T}{\partial s} s' + \frac{\partial T}{\partial \rho} \rho' + \frac{\partial T}{\partial w_{jl}} w_{jl}'.
\end{align}
Since $\partial T/\partial w_{jl} = \partial \lambda_{jl}/\partial s$, and $K$ and $\mu_V$ depend only weakly on $s$, by \eqref{lambda} we neglect $ {\partial T}/{\partial w_{jl}} $.  Substitution for $\rho'$ and $s'$ from \eqref{rho2} and \eqref{s2} then yields
\begin{align}
T' \approx s \frac{\partial T}{\partial s} \frac{ k_i }{\omega} {v_n}_i' + \frac{\partial T}{\partial \rho} \frac{k_i}{\omega} g_i'.
\end{align}
We earlier showed that the mode is longitudinal, so we drop the indices $i$.  
The identity $g' = \rho \left( 1+ {j_s'}/{j_n'} \right) v_n'$ then yields, for both modes, that 
\begin{align}
T' \approx \frac{ k }{\omega} {v'_n}\left[s \frac{\partial T}{\partial s}  + \rho \frac{\partial T}{\partial \rho}  \left( 1+ \frac{j_s'}{j_n'} \right)\right] .
\label{Tgeneral}
\end{align}

Substituting the ratio ${j_s'}/{j_n'}$ from \eqref{js1TOjn1-2} and \eqref{js4TOjn4-2} and $\omega_{1,4}=+ck_{1,4}$ from \eqref{FirstSoundFreq-2} and \eqref{FourthSoundFreq-2} into \eqref{Tgeneral} gives the temperature associated with each mode.  To lowest order in $f_s$ we obtain
\begin{align}
T'_1 \approx& \frac{ {v_n}_1' }{c_1}  \left[s \frac{\partial T}{\partial s}  + \rho \frac{\partial T}{\partial \rho}  \right] \label{T10},\\
T'_4 \approx& f_s^{-\frac{1}{2}} \frac{{v_n}_4'}{\widetilde{c}_0}
 \left[s \frac{\partial T}{\partial s}  + \rho \frac{\partial T}{\partial \rho}  \left(1 - \frac{c_1^2}{\widetilde{c}^2 } \right) \right] \label{T40}.
\end{align}
For a solid at low temperature, Appendix~\ref{dTAppendix} gives $\rho (\partial T/\partial \rho) \approx 10 s (\partial T/\partial s)$.  In addition, for $P_a \ll K$ (as is the case here), Appendix~\ref{ApproximationAppendix} gives $c_1^2 \gg \widetilde{c}^2$.  Therefore \eqref{T10}-\eqref{T40} become
\begin{align}
T'_1 \approx& \rho  \frac{\partial T}{\partial \rho} \frac{ {v_n}_1' }{c_1}  \label{T1},\\
T'_4 \approx& -f_s^{-\frac{1}{2}} \rho \frac{\partial T}{\partial \rho}  \frac{c_1^2}{\widetilde{c}^2} \frac{{v_n}_4'}{\widetilde{c}_0}= -f_s^{-\frac{1}{2}}  \rho \frac{\partial T}{\partial \rho}\frac{c_1^2}{\widetilde{c}^2}\frac{{v_n}_4'}{{v_n}_1'}\frac{{v_n}_1'}{\widetilde{c}_0} . \label{T4}
\end{align}

The total temperature deviation therefore is
\begin{align}
T' = T_1' + T_4' \approx  \rho  \frac{\partial T}{\partial \rho}  \frac{ {v_n}_1' }{c_1}   \left[1  -f_s^{-\frac{1}{2}}  \frac{c_1^3}{\widetilde{c}^2 \widetilde{c}_0} \frac{ {v_n}_4'}{{v_n}_1'} \right].
\label{Ttot}
\end{align}
The ratio ${v_n'}_4/{v_n'}_1$ depends on the mode generator, to be discussed in the next section.

\section{Longitudinal Mode Generation}
\label{ModeDetectionSection}
A transducer produces, and therefore can be used to detect, stress deviations.  A heater produces, and therefore can be used to detect, temperature deviations (when used as a detector, a heater is called a thermometer).  To utilize the results of Sec.~\ref{StressTemp}, we find ${v_n'}_4/{v_n'}_1$ for each device, then substitute it into \eqref{sigmaFirst}-\eqref{sigmatot} and \eqref{T1}-\eqref{Ttot} to find the respective stress and temperature deviations produced by transducers and heaters.

\subsection{Transducer Properties}
For a transducer we take $v_s' = v_n'$ (and therefore $j_s'=0$) so that
\begin{align}
0 = j_s'|_{\rm trn} = \left[{j_s'}_1 + {j_s'}_4 \right]_{\rm trn} = \left[ \frac{{j_s'}_1}{{j_n'}_1} {j_n'}_1 + \frac{{j_s'}_4}{{j_n'}_4} {j_n'}_4 \right]_{\rm trn},
\end{align}
where the subscript ``trn'' denotes properties of a transducer.
Then 
\begin{align}
\left. \frac{{v_n'}_4}{{v_n'}_1} \right|_{\rm trn} = \left. \frac{{j_n'}_4}{{j_n'}_1} \right|_{\rm trn}= - \frac{{j_s'}_1/{j_n'}_1}{{j_s'}_4/{j_n'}_4}.
\end{align}
Use of \eqref{js1TOjn1-2} and \eqref{js4TOjn4-2} yields
\begin{align}
\left. \frac{{v_n'}_4}{{v_n'}_1} \right|_{\rm trn}  \approx& -f_s\frac{\widetilde{c}^2}{c_1^2} \left(1 - \frac{\widetilde{c}^2}{c_1^2}  \right) \approx -\frac{f_s \widetilde{c}^2}{c_1^2},
\label{vn4TOvn1TRANS}
\end{align}
where we have taken $c_1^2 \gg \widetilde{c}^2 $ (see Appendix~\ref{ApproximationAppendix}).

Eq.~\eqref{sigmaFirst} gives $\hat{\sigma}'_1$ in terms of ${v_n'}_1$, regardless of generator.  Use of \eqref{vn4TOvn1TRANS} in \eqref{sigmaFourth}-\eqref{sigmatot} gives 
\begin{align}
\left. \hat{\sigma}'_4 \right|_{\rm trn} \approx&  - f_s^{\frac{3}{2}} \rho \widetilde{c}_0 {v_n'}_1, \label{sigmaFourthTRANS}\\
\left. \hat{\sigma}'\right|_{\rm trn} \approx& - \rho c_1 {v_n'}_1 \left( 1 + f_s^{ \frac{3}{2}} \frac{\widetilde{c}_0}{c_1}\right) \approx  - \rho c_1 {v_n'}_1,
\label{sigmatotTRANS}
\end{align}
for $f_s \ll 1$ and $\widetilde{c}_0^2 \ll c_1^2$. 
Thus the stress produced by a transducer primarily goes into first sound, with a fraction $f_{s}^{3/2}(\widetilde{c}_0/c_{1})$ of the stress going into fourth sound.  Eqs.~\eqref{sigmaFirst} and \eqref{sigmaFourthTRANS} divided by \eqref{sigmatotTRANS} are the two entries in the top left of Table~\ref{EfficiencyTable}.

Eq.~\eqref{T1} gives $T'_1$ in terms of ${v_n'}_1$, regardless of generator.   Substituting \eqref{vn4TOvn1TRANS} into \eqref{T4} yields
\begin{align}
\left. T'_4 \right|_{\rm trn} \approx& f_s^{\frac{1}{2}}  \rho  \frac{\partial T}{\partial \rho} \frac{{v_n}_1'}{\widetilde{c}_0}  \label{T4TRANS}.
\end{align}
By \eqref{T1}, $T_4'=f_s^{1/2}(c_1/\widetilde{c}_0)T_1'$, so $T_1'$ and $T_4'$ could be of the same order of magnitude.   Eqs.~\eqref{T1} and \eqref{T4TRANS} divided by \eqref{sigmatotTRANS} are the two entries in the top right of Table~\ref{EfficiencyTable}. 

\subsection{Heater Properties}
For a heater we take $g' = 0 $, so that
\begin{align}
0 =& g'|_{\rm htr} = \left[({j_s'}_1 + {j_n'}_1) + ({j_s'}_4 + {j_n'}_4) \right]_{\rm htr}\notag\\
=& \left(\frac{{j_s'}_1}{{j_n'}_1} + 1 \right) {j_n'}_1 |_{\rm htr} + \left(\frac{{j_s'}_4}{{j_n'}_4} + 1 \right) {j_n'}_4 |_{\rm htr},
\end{align}
where the subscript ``htr'' denotes properties of a heater.
Then 
\begin{align}
\left. \frac{{v_n'}_4}{{v_n'}_1} \right|_{\rm htr} = \left. \frac{{j_n'}_4}{{j_n'}_1} \right|_{\rm htr} =  - \frac{({j_s'}_1/{j_n'}_1) + 1 }{({j_s'}_4/{j_n'}_4) + 1},
\end{align}
and substitution from \eqref{js1TOjn1-2} and \eqref{js4TOjn4-2} yields, for $f_s \ll 1$,
\begin{align}
\left. \frac{{v_n'}_4}{{v_n'}_1} \right|_{\rm htr} \approx&  \frac{1}{\frac{c_1^2}{\widetilde{c}^2} - 1} \approx  \frac{\widetilde{c}^2}{c_1^2}.
\label{vn4TOvn1HEAT}
\end{align}
Here, we have used $c_1^2 \gg \widetilde{c}^2$ (see Appendix~\ref{ApproximationAppendix}).

Eq.~\eqref{T1} gives $T'_1$ in terms of ${v_n'}_1$, regardless of generator. 
Substitution of \eqref{vn4TOvn1HEAT} into \eqref{T4}-\eqref{Ttot} gives 
\begin{align}
\left. T'_4 \right|_{\rm htr}  \approx& -f_s^{-\frac{1}{2}}  \rho  \frac{\partial T}{\partial \rho} \frac{{v_n}_1'}{\widetilde{c}_0} \label{T4HEAT},\\
\left. T' \right|_{\rm htr} \approx&  \rho  \frac{\partial T}{\partial \rho} \frac{ {v_n}_1' }{c_1} \left[1  -f_s^{-\frac{1}{2}} \frac{c_1}{\widetilde{c}_0}  \right] \approx  -  f_s^{-\frac{1}{2}}  \rho \frac{\partial T}{\partial \rho} \frac{ {v_n}_1'}{\widetilde{c}_0} ,
\label{TtotHEAT}
\end{align}
for $f_s \ll 1$ and $\widetilde{c}_0^2 \ll c_1^2$.
Thus the temperature produced by a heater primarily goes into fourth sound, with a fraction $f_{s}^{1/2}(\widetilde{c}_0/c_{1})$ of the temperature going into first sound.  Eqs.~\eqref{T1} and \eqref{T4HEAT} divided by \eqref{TtotHEAT} are the two entries in the bottom right of Table~\ref{EfficiencyTable}.

Eq.~\eqref{sigmaFirst} gives $\hat{\sigma}'_1$ in terms of ${v_n'}_1$, regardless of generator.  
Substituting \eqref{vn4TOvn1HEAT} into \eqref{sigmaFourth} yields 
\begin{align}
\left. \hat{\sigma}'_4 \right|_{\rm htr} \approx&   f_s^{\frac{1}{2}} \rho \widetilde{c}_0 {v_n'}_1. \label{sigmaFourthHEAT}
\end{align}
Since $f_s \ll 1$ and $\widetilde{c}_0^2 \ll c_1^2$, we have $\hat{\sigma}'_1 \gg \hat{\sigma}'_4$.   Eqs.~\eqref{sigmaFirst} and \eqref{sigmaFourthHEAT} divided by \eqref{TtotHEAT} are the two entries in the bottom left of Table~\ref{EfficiencyTable}.

\subsection{Generation Efficiencies}

A proper treatment of the response of a given detector (transducer or thermometer) to a given mode (first or fourth sound) would consider the incoming mode and what happens under reflection from the detector; this would give the net stress and temperature at the detector.\cite{Khalatnikov}  We consider only the issue of generation.

A transducer generates mostly stress, $\hat{\sigma}'$.  Thinking of the equation entries in Table~\ref{EfficiencyTable} as a 4-by-2 matrix $\cal{M}$, ${\cal M}_{11}$ shows that a transducer is efficient as a first sound generator. 
${\cal M}_{21}/{\cal M}_{11}$ gives 
\begin{align}
\frac{\hat{\sigma}'_4}{\hat{\sigma}'_1} \approx f_s^{3/2} \frac{\widetilde{c}_0}{c_1}.    \qquad {\rm (transducer)}
\label{sigma4TOsigma1TRANS}
\end{align}
For $f_s \ll 1$ and $\widetilde{c}_0^2 \ll c_1^2$, this is negligible.

Although a transducer primarily produces stress, it also produces a small temperature deviation $T'$.  ${\cal  M}_{22}/{\cal  M}_{12}$ gives 
\begin{align}
\frac{T'_4}{T'_1} \approx f_s^{1/2} \frac{c_1}{\widetilde{c}_0}.     \qquad {\rm (transducer)}
\label{T4TOT1TRANS}
\end{align}
For $f_s \ll 1$ and $\widetilde{c}_0^2 \ll c_1^2$, it is not clear which of the terms in \eqref{T4TOT1TRANS} dominates.  

A heater generates mostly temperature, $T'$, and ${\cal M}_{42}$ shows that a heater is efficient as a fourth sound generator.  ${\cal  M}_{42}/{\cal  M}_{32}$ gives 
\begin{align}
\left| \frac{T'_4}{T_1'} \right| \approx f_s^{-{1}/{2}} \frac{c_1}{\widetilde{c}_0} \gg 1,    \qquad {\rm (heater)}
\label{T4TOT1HEAT}
\end{align}  
which is large.  In fact, for $f_s \ll 1$ and $\widetilde{c}_0^2 \ll c_1^2$, ${\cal M}_{42} = {T'_4}/{T'} \approx 1$ and therefore nearly all of the temperature response corresponds to the fourth sound mode. 
 
Although a heater primarily produces temperature, it also produces a small stress deviation $\hat{\sigma}'$.  ${\cal M}_{41}/{\cal M}_{31}$ gives 
\begin{align}
\left| \frac{\hat{\sigma}'_4}{\hat{\sigma}'_1} \right| \approx f_s^{1/2} \frac{\widetilde{c}_0}{c_1}.  \qquad {\rm (heater)}
\label{sigma4TOsigma1HEAT}
\end{align}
Therefore, with $f_s \ll 1$ and $\widetilde{c}_0^2 \ll c_1^2$, 
eqs.~\eqref{sigma4TOsigma1TRANS} and \eqref{sigma4TOsigma1HEAT} imply that stress deviations do not contribute an appreciable amount of fourth sound, whether produced by a transducer or a heater.

\begin{table*}[floatfix]
\begin{tabular}{c | c |c |c }
\hline \hline
&&&\\
Generator & Mode & Stress & Temperature \\[2ex] \cline{3-4} \hline
\multicolumn{1}{c|}{\,\,\,\, Transducer\,\,\,\,}&\,\,\,\, 1st Sound \,\,\,\,& $\displaystyle \frac{\hat{\sigma}'_1}{ \hat{\sigma}'} \approx \frac{ 1}{\displaystyle 1 + f_s^{\frac{3}{2}} \frac{\widetilde{c}_0}{c_1}} \approx 1$ & $\displaystyle \frac{ T_1'}{ \hat{\sigma}'} \approx -\frac{1}{c_1^2} \frac{\partial T}{\partial \rho} $ \\ [5ex] \cline{2-4} 
\multicolumn{1}{c|}{ }&4th Sound & $\displaystyle \frac{ \hat{\sigma}'_4}{ \hat{\sigma}'} \approx \frac{ 1}{\displaystyle 1 + f_s^{-\frac{3}{2}} \frac{c_1}{\widetilde{c}_0}} \approx f_s^{\frac{3}{2}} \frac{\widetilde{c}_0}{c_1} \ll 1$ & $\displaystyle \frac{ T_4'}{ \hat{\sigma}'} \approx -\frac{f_s^{\frac{1}{2}}}{c_1\widetilde{c}_0} \frac{\partial T}{\partial \rho} = \frac{f_s^{\frac{1}{2}} c_1}{\widetilde{c}_0} \frac{ T_1'}{ \hat{\sigma}'}$ \\ [5ex] \hline 
\multicolumn{1}{c|}{Heater} &1st Sound& $\displaystyle \frac{ \hat{\sigma}'_1}{ T'} \approx f_s^{\frac{1}{2}} c_1 \widetilde{c}_0 \frac{\partial \rho}{\partial T}$ &\,\,\,\, $\displaystyle \frac{ T_1'}{ T'} \approx \frac{\displaystyle 1}{\displaystyle 1 - f_s^{-\frac{1}{2}} \frac{c_1}{\widetilde{c}_0}} \approx -f_s^{\frac{1}{2}} \frac{\widetilde{c}_0}{c_1} \ll 1$ \,\,\,\, \\ [5ex] \cline{2-4}
\multicolumn{1}{c|}{ }&4th Sound&\,\,\,\, $\displaystyle \frac{ \hat{\sigma}'_4}{ T'} \approx f_s \widetilde{c}_0^2 \frac{\partial \rho}{\partial T} = -f_s^{\frac{1}{2}} \frac{\widetilde{c}_0}{c_1} \frac{\hat{\sigma}'_1}{ T'} \ll \frac{\hat{\sigma}'_1}{ T'}$ \,\,\,\,& $\displaystyle \frac{ T_4'}{ T'} \approx \frac{\displaystyle 1}{\displaystyle 1 - f_s^{\frac{1}{2}} \frac{\widetilde{c}_0}{c_1}} \approx 1$ \\ [5ex]
\hline \hline
\end{tabular}
\caption{The efficiency of first and fourth sound mode generation by transducers and heaters, with entries given as a 4-by-2 matrix ${\cal M}$.  $T'$ is the temperature deviation (produced/detected by a heater/thermometer) and $\hat{\sigma}'=\sigma'_{11}$ is the longitudinal stress deviation (produced and detected by a transducer), where subscripts $1$ and $4$ denote the first and fourth sound modes.  Here, $ \partial T/\partial \rho$ (and $ \partial \rho/\partial T$) is taken at constant $s$ and $w_{jl}$. }
\label{EfficiencyTable}
\end{table*}

\section{Summary}
\label{Conclusion}
We have studied the implications of the Andreev and Lifshitz theory of supersolids for the generation of a fourth sound mode in a solid under an applied pressure $P_a$, including the relative efficiencies of a transducer and a heater in producing both longitudinal elastic waves and fourth sound waves.  The present results apply when the bulk modulus $K \gg P_a$.

\section{Acknowledgements}
We would like to thank H. Kojima for prompting us to undertake this calculation.  We acknowledge the support of the Department of Energy under grant DE-FG02-06ER46278.

\appendix

\section{Relating Applied Pressure $P_{a}$ and Internal Pressure $P$}
\label{PPaAppendix}
We now use the experimental data of Ref.~\onlinecite{FranckWanner} to estimate $P/P_a$.  We then evaluate when $P_a \gg P$. 

 Ref.~\onlinecite{SearsSasOrdinary} gives that 
 \begin{align}
 \frac{P}{P_a} \approx \frac{K^{{}^*} P_{a}}{2K^2},
\label{P/Pa}
\end{align}
where $K^* \equiv K - V (\partial K/\partial V)|_{ w_{ik},\sigma,N}$.  
Under hydrostatic compression $\lambda_{ik}^{(0)} - P \delta_{ik} = -P_a \delta_{ik}$.  Thus,
\begin{align}
\lambda_{ll}^{(0)} = 3P - 3P_a\approx \frac{3 K^{{}^*} P_{a}^2}{2K^2} - 3 P_a.
\label{lambdaTOPPa}
\end{align}

Unfortunately, $K^*$ is not a quantity measured experimentally, since the structure of the energy density 
dictates that the derivative is taken at constant strain (i.e., constant lattice site density).

In what follows, we roughly estimate $\partial K/\partial V$ by assuming it to be of the same order of magnitude whether taken at constant $w_{ik}$ or under typical experimental conditions.  That is, we take
\begin{align}
\left. \frac{\partial K}{\partial V} \right|_{w_{ik},\sigma,N} \approx \left. \frac{\partial K}{\partial V} \right|_{\sigma,N,\rm exp}\approx \left. \frac{\Delta K}{\Delta V} \right|_{\sigma,N,\rm exp}.
\end{align}

We now consider the data of Ref.~\onlinecite{FranckWanner}.  Although the samples were necessarily under pressure, Ref.~\onlinecite{FranckWanner} appears to apply $c_1^2 = (1/\rho)(\partial \lambda/\partial w) =(1/\rho)[K+(4/3)\mu_V]$ without including corrections due to $P_{a}$.\cite{SearsSasOrdinary}   
Nevertheless their result should permit a rough estimate (for simplicity we consider that $T\approx 0$).  We use $K = (1/3)[c_{11} + 2c_{13}]$, where $c_{11}$ and $c_{13}$ are elastic constants.\cite{FranckWanner}. 
Select parts of Tables~I and II of Ref.~\onlinecite{FranckWanner} are reproduced in Table~\ref{FWTable} for two molar volumes, which is sufficient to make estimates.

These data give $\Delta V \approx -1.2 $~cm$^3$/mole and $\Delta K  \approx 160$~bars, so that $\Delta K/\Delta V \approx -133$~bars-mole/cm$^3$, which we take to be constant since the elastic constants in Figure~I of Ref.~\onlinecite{FranckWanner} are linear in volume.   
Thus 
we obtain the two values for $K^*$ in 
Table~\ref{FWTable}: 
$K^* \approx 10 K$ at $P_a = 31$~bars and $K^* \approx 6.6 K$ at $P_a = 52$~bars.

\begin{table*}[floatfix]
\begin{tabular}{ c  c  c  c  c  c  c  c  c  c }
\hline \hline
&&&&&&&&&\\
Volume \,\,\,\,& \,\,\,$P_a$ \,\,\,& $c_{33}$ & $c_{13}$ & $K$ & $\Delta K/\Delta V$   & $K^*$  & $P$  &\,\,\,\, $P/P_a$ \,\,\,\,&\,\,\,\, $P_a/K$ \\  [1ex]
$\left({\rm cm}^3/\rm mole\right)$ \,\,\,&\,\,\, (bars) \,\,\,&\,\,\, (bars) \,\,\,&\,\,\,  (bars)  \,\,\,& \,\,\, (bars) \,\,\,& \,\,\, $\left({\rm bars \,\, mole}/{\rm cm}^3 \right)$ \,\,\,&\,\,\, (bars) \,\,\,\,&\,\,\,\, (bars) \,\,\,\,&  &  \\ [2ex]
 \hline 
19.28 & 51.6$^{\dagger}$ & 980 & 198 & 460  & -133 & 3020 & 19.1 & 0.37 &  0.11\\ [1ex]
20.5 & 31.4$^{\dagger}$ & 630 & 142 & 300 & -133 & 3030 & 16.6 & 0.53 & 0.10 \\ \hline \hline

\end{tabular}

\caption{The first four columns are experimental data on hcp $^4$He from Ref.~\onlinecite{FranckWanner}.  We also employ $K=[c_{33} + 2 c_{13}]/3$, assume $K$ to be linear in $V$, take $K^* \approx K - V (\Delta K/\Delta V)$, and find $P/P_a$ from \eqref{P/Pa}. $^\dagger$Value was estimated by extrapolation from Table I of Ref.~\onlinecite{FranckWanner}.  }
\label{FWTable}

\end{table*}

For $P_a \approx 31 $~bars, $P/P_{a}\approx0.53$ (and thus $\lambda_{11}/P_{a}\approx0.47$). 
For $P_a \approx 52 $~bars, $P/P_{a}\approx0.37$ (and thus $\lambda_{11}/P_{a}\approx0.63$).


For $V=20.5$~cm$^3$/mole,  $P_a/K \approx 0.10 $, so it is appropriate to take $P_a \ll K$.  For $V=19.28$~cm$^3$/mole,  $P_a/K \approx 0.11 $, and $P_a \ll K$ is still a reasonable approximation.  Therefore, for applied pressures less than 100 bars (and possibly higher), $P_a \ll K$ likely holds.

$P_{a}/K$ increases as $P_{a}$ increases.  Therefore, although at higher $P_{a}$ we may find that $P_a \gg P$ (extrapolating from Table~\ref{FWTable}), $P_a/K$ might become on the order of unity, and the approximations made in the present work and in Ref.~\onlinecite{SearsSasOrdinary} no longer apply.


\section{Velocities, Thermodynamic Derivatives and Strain in a Crystal Under Applied Pressure}
\label{ApproximationAppendix}
We now estimate the relative sizes of the velocities $c_1$, $c_0$, $\widetilde{c}$, and $\widetilde{c}_0$ in the limit $P_a \ll K$.  We use the relationships between thermodynamic derivatives and applied pressure given in Ref.~\onlinecite{SearsSasOrdinary}.

To lowest order in $P_a/K$, Ref.~\onlinecite{SearsSasOrdinary} gives 
\begin{align}
&w_{ll}^{(0)} = - \frac{P_a}{K},\label{wllAppend}\\
&\frac{\partial \lambda}{\partial \rho} =  \frac{V P_a}{\rho K} \left. \frac{\partial K}{\partial V}\right|_{\sigma, w_{ik}, N} ,\label{dlambdadrhoAppend}\\
&\frac{\partial P}{\partial \rho} = \frac{V^2 P_a^2}{2 \rho K^2} \left. \frac{\partial ^2 K}{\partial V^2} \right|_{\sigma, w_{ik}, N},\label{dPdrhoAppend}\\
&\frac{\partial P}{\partial w} =  -P_a \left(1 -  \frac{V}{K} \left. \frac{\partial K}{\partial V}\right|_{\sigma, w_{ik}, N} \right),\label{dPdwAppend}\\
&c_1^2 \approx \frac{K+\frac{4}{3} \mu_V}{\rho}, \label{c1Append}
\end{align}
where $\sigma = s/\rho$.  Here the internal pressure $P$ has been taken to depend {\it only} on the square of the strain.  
Although $\partial \lambda/\partial \rho$ and $\partial P/\partial \rho$ in Ref.~\onlinecite{SearsSasOrdinary} are taken at constant $\sigma$, not $s$, at solid $^{4}$He temperatures we assume that $\sigma\approx0\approx s$.  Note that 
to lowest order in $P_a/K$, the strain of eq.~\eqref{wllAppend} agrees with Ref.~\onlinecite{LLElasticity}, which includes lattice stress but no internal pressure $P$.  (Ref.~\onlinecite{SearsSasOrdinary} also finds a $P_{a}^{2}$ term in the strain that is not obtained in Ref.~\onlinecite{LLElasticity}.) 
 
We now use the Gibbs-Duhem relation \eqref{GibbsDuhem} to determine $c_0^2$:
\begin{align}
c_0^2 = \rho \frac{\partial \mu}{\partial \rho} \approx& \frac{\partial P}{\partial \rho} - w_{ll}^{(0)} \frac{\partial \lambda}{\partial \rho} \notag\\
\approx& \frac{V P_a^2}{\rho K^2}\left[\frac{V}{2} \left. \frac{\partial ^2 K}{\partial V^2} \right|_{\sigma, w_{ik}, N} +  \left. \frac{\partial K}{\partial V}\right|_{\sigma, w_{ik}, N} \right].
\label{c0Append}
\end{align}
Note that Ref.~\onlinecite{AL69} takes $\rho (\partial \mu/\partial \rho) = \partial P/\partial \rho$, and thus does not include the term proportional to the static strain. 
As for $P$, $\mu$ depends {\it only} on the square of the strain, via a Maxwell relation.  This is not true for a good liquid.  
 
Eq.~\eqref{c0Append} shows that $c_0^2$ is second order in $P_a/K$, whereas \eqref{dlambdadrhoAppend} shows that $\partial \lambda/\partial \rho$ is first order in $P_a/K$.  Thus, for $P_a \ll K$, $|\partial \lambda/\partial \rho| \gg |c_0^2|$.  Therefore \eqref{gammaDEF} gives 
 \begin{align}
 \widetilde{c}^2 \approx -\partial \lambda/\partial \rho \gg c_0^2.
\label{tildecTOc0}
 \end{align}

Further, we may find the sign of $\widetilde{c}^2$.  $K$ is a measure of the stiffness of a solid.  Thus, as $V$ increases at constant particle number and strain (or lattice site number density), i.e., as vacancies and lattice sites are added to the system, $K$ should decrease, or $\partial K/\partial V < 0$.  Then \eqref{dlambdadrhoAppend} gives $\partial \lambda/\partial \rho <0$, so that 
\begin{align}
\widetilde{c}^2 \approx -\partial \lambda/\partial \rho > 0.
\label{tildecpositive}
\end{align}

Moreover, since to first order $c_1^2$ is independent of $P_a$, eqs.~\eqref{dlambdadrhoAppend} and \eqref{c1Append} give 
\begin{align}
c_1^2 \gg |\partial \lambda/\partial \rho|. 
\label{c1TOdlambdadrho}
\end{align}   
Combining \eqref{tildecTOc0} and \eqref{c1TOdlambdadrho} yields
\begin{align}
c_1^2 \gg \widetilde{c}^2 \gg c_0^2. 
\label{c1TOtildecTOc0}
\end{align}   

Finally, \eqref{barc4} gives $\widetilde{c}_0^2 \approx c_0^2 - (\widetilde{c}^4/c_1^2)$, which implies that $\widetilde{c}_0^2 \ll \widetilde{c}^2$.  Thus, \eqref{c1TOtildecTOc0} gives
\begin{align}
c_1^2 \gg \widetilde{c}^2 \gg \widetilde{c}_0^2.
\label{barc4llc1}
\end{align}
The quantities $c_0^2$ and $\widetilde{c}_0^2$ may be of the same order.  

\section{On $s(\partial T/\partial s)$ and $\rho (\partial T/\partial \rho)$}
\label{dTAppendix}

For an insulating solid at low temperatures, it is well-known that $s = \xi (T/\overline{u})^{3}$, where $\xi$ is a dimensionless constant and $\overline{u}$ is the mean velocity of longitudinal and transverse ordinary sound.\cite{LLStatistical}  It immediately follows that 
\begin{align}
s \left.\frac{\partial T}{\partial s}\right|_{\rho} = \frac{T}{3}, 
\label{sdTds}
\end{align}
and that $(\partial T/\partial\overline{u})_{s}=T/\overline{u}$.  We then have 
\begin{align}
\rho\frac{\partial T}{\partial \rho}\Big|_s = \rho\frac{\partial T}{\partial \overline{u} }\Big|_s \frac{\partial \overline{u} }{\partial \rho}\Big|_{s}\approx T \frac{\rho}{\overline{u} } \frac{\partial \overline{u} }{\partial \rho}|_{s}.
\label{rhodTdrho}
\end{align}
We are at low enough temperatures that we may consider $\overline{u}$ to depend only on density, so that both $s$ and $T$ may be considered nearly zero.

We now estimate $({\rho}/{\overline{u}})({\partial \overline{u}}/{\partial \rho})|_{T}=-({\cal V}/{\overline{u}})({\partial \overline{u}}/{\partial {\cal V}})|_{T}$ of \eqref{rhodTdrho}, where the molar volume ${\cal V}\sim\rho^{-1}$.  Using data from Fig.13 of Ref.~\onlinecite{Greywall} we take an averaged longitudinal sound velocity of $5\times10^{4}$~cm/s to be $\overline{u}$, $\cal{ V}$ to be $20$~cm$^3$/mole, and (from Fig.13) $\Delta\overline{u}/\Delta{\cal V}\approx(\partial \overline{u}/\partial{\cal V})$ to be $-0.83\times10^{4}$~mole/(cm$^{2}$-s).  This gives $({\rho}/{\overline{u}})({\partial \overline{u}}/{\partial \rho})|_{T}\approx3.3$.  Comparison with \eqref{sdTds} yields $ \rho (\partial T/\partial \rho) \approx 10\, s (\partial T/\partial s) $.


\begin{thebibliography}{}
\bibitem{Thouless69} D. J. Thouless, Ann. Phys. (N.Y.) 52, 403 (1969).  This contains the remark that, for a lattice of bosons, vacancies could be ``in the lowest Bloch state with a finite probability, so the system would be `super' but not `fluid'\,''. 
\bibitem{GalileanNote} Sect.V of Ref.~\onlinecite{Saslow77} takes $\vec{v}_{s}$ to be defined solely relative to the lattice, and thus $\vec{v}_{s}$ was not taken to be a Galilean velocity.  Therefore the normal mode results in that work do not apply to the present system.
\end{thebibliography}
\end{document}